\documentclass[a4paper]{jpconf}
\usepackage{graphicx}
\begin{document}
\title{The Muon g-2 Experiment Overview and Status as of June 2016}
\author{J. L. Holzbauer}
\address{University of Mississippi, University, Mississippi 38677, USA}
\ead{jholzbau@fnal.gov}

%NOTES:
%need to fix e34 citation to include CDR not just author...

\begin{abstract}
%abstract is partially taken from the conference talk abstract submitted in April 2016
The Muon g-2 Experiment at Fermilab will measure the anomalous magnetic moment of the muon to a precision of 140 parts per billion, which is a factor of four improvement over the previous E821 measurement at Brookhaven. The experiment will also extend the search for the electric dipole moment (EDM) of the muon by approximately two orders of magnitude, with a sensitivity down to $10^{-21}$ e.cm. Both of these measurements are made by combining a precise measurement of the 1.45T storage ring magnetic field with an analysis of the modulation of the decay rate of higher-energy positrons (from anti-muons), recorded by 24 calorimeters and 3 straw tracking detectors. The recent progress in the alignment of the electrostatic quadrapole plates and the trolley rails inside the vacuum chambers, and in establishing the uniform storage ring magnetic field will be described.
\end{abstract}

\section{Introduction}
In 2006, an experiment at BNL called E821 reported a final, precisely measured value of g-2 that was approximately 3 sigma away from the theoretical value~\cite{e821}.  In recent years, this tension has remained despite improvements in the theoretical calculation~\cite{tdr}.  To help resolve this tension, two new experiments are being built or planned.  One is E34, an ultra-cold muon beam experiment in Japan~\cite{e34, e34proc}.  The other is an improved version of E821, the Muon g-2 Experiment~\cite{proposal, tdr}, which has reference number E989 and is located at Fermilab in the US.  In this proceedings, we will discuss the physics of g-2 and the current standard model (SM) value, the Muon g-2 Experiment procedure and setup, and progress in assembling the new experiment.

\section{Standard Model and Current Experimental Value}
The quantity g is a constant which relates particle spin and magnetic moment.  In Dirac theory, the value of g is exactly 2~\cite{dirac}.  However, higher order effects will alter this value.  For example, new physics which interacts weakly, such as a new heavy W, could contribute in this way in SUSY models~\cite{snowmass}.  The muon is studied rather than the electron in this experiment because the muon is much heavier and thus is more sensitive to higher order corrections but still long lived enough to study.  The experiment specifically studies what is known as the anomaly, which is $a_{\mu}=$(g-2)/2.

The standard model theory calculation breaks down into four main components: quantum electrodynamics (QED)~\cite{smqed}, electroweak (EW)~\cite{smew}, hadronic vacuum polarization (HVP)~\cite{smhpv1,smhpv2} and hadronic light-by-light (HLbL)~\cite{smhlbl}.  The first two, the QED and EW components are well known with relatively small uncertainties.  As can be seen in Table~\ref{SM}, the QED component is the largest contribution to the standard model value but has the smallest uncertainty, while the hadronic terms are less well known. The contribution from hadronic vacuum polarization can be obtained from e+e- to hadrons or tau to hadrons measurements.  The hadronic light-by-light term cannot be obtained directly from data, though data can be used indirectly to constrain the contribution.  Rather, the value is obtained diagram or lattice calculations.

\begin{table}[h]
\caption{\label{SM} Standard model components of the anomaly, taken directly from~\cite{tdr}.  Two values are shown for HVP to reflect two recent estimates.  The terms lo and ho indicate lower order and higher order, respectively.  Other terms are defined in the text.}
\begin{center}
\begin{tabular}{lr}
\br
&Values in $10^{-11}$ units\\
\mr
QED ($\gamma$ + $l$) & 116584718.951 $\pm$ 0.009 $\pm$ 0.019 $\pm$ 0.007 $\pm$ 0.077\\
HVP(lo)~\cite{smhpv1} & 6923 $\pm$ 42\\
HVP(lo)~\cite{smhpv2} & 6949 $\pm$ 43\\

HVP(ho)~\cite{smhpv2} & -98.4 $\pm$ 0.7\\
HLbL         & 105 $\pm$ 26\\
EW           & 153.6 $\pm$ 1.0\\
\mr
Total SM~\cite{smhpv1} & $116591802 \pm 42_{\rm{H-LO}} \pm 26_{\rm{H-HO}} \pm 2_{\rm{other}}$ ( $\pm 49_{\rm{tot}}$)\\
Total SM~\cite{smhpv2} & $116591828 \pm 43_{\rm{H-LO}} \pm 26_{\rm{H-HO}} \pm 2_{\rm{other}}$ ( $\pm 50_{\rm{tot}}$)\\
\br
\end{tabular}
\end{center}
\end{table}

The value of the anomaly from the BNL experiment~\cite{e821}, corrected for updated constants~\cite{codata} and given in~\cite{tdr}, is $a_{\mu}^{E821} = 116 592 089 \pm 63 \times 10^{-11}$ (54~ppm).  The new experiment is expected to reduce this uncertainty by a factor of four to 0.14~ppm, which would give a 5 sigma deviation from the standard model calculation, assuming the central value remains the same for both the standard model calculation and the measurement.  New data inputs and improvements to the lattice calculations are expected to reduce the hadronic contribution uncertainties (particularly HVP lo), which are the largest sources of theoretical uncertainty.  If the standard model uncertainty improves in addition to the experimental uncertainty, the deviation could be around 8 sigma under these same central value assumptions~\cite{tdr}.

\section{The Muon g-2 Experiment Plans and Status}
The Muon g-2 Experiment is located at Fermilab.  It reuses the BNL storage ring, which was moved carefully from BNL to Fermilab.  Most parts, like vacuum chambers, could be shipped by normal means, but the 15 ton cryostat ring was trickier.  It is large, heavy and the superconducting coils cannot flex more than approximately 3~mm.  This portion was moved by the Emmert Corportation, which used a barge and a large truck.  In 2015, the magnet and related cryo-systems were tested, and the magnet was cooled and powered at Fermilab.  The required 1.45 Tesla field was achieved and the transportation was deemed a success.

The muons for the storage ring are produced by the Fermilab accelerator complex.  The source protons are produced and moved by reused Tevatron equipment, and they collide with a lithium target to produce pions.  These pions enter a long delivery ring where they decay to muons and neutrinos.  The polarized, positively charged muon beam is then injected into the storage ring.  The long delivery ring helps to reduce proton/pion contamination, which was an issue at BNL.  The new experiment expects a factor of 20 increase in muon statistics over the BNL experiment and which will reduce the statistical error to 0.1~ppm.

In brief, the experiment consists of a cryo-system surrounding dipole magnets, which contain vacuum chambers with electrostatic quadrapole plates.  The magnets and quadrapole plates work in tandem to keep the muons in the proper location.  There are other beam related parts like the inflector, kickers, collimators and a beam monitoring system, as well as a trolley that rides around inside of the ring to measure the field.  Detector systems include both straw trackers and lead crystal calorimeters.  Trackers were not used in the BNL experiment and should add valuable information to the new g-2 data.  Additionally, the calorimeters now have multiple read out channels to reduce pileup.

The measurement consists of two quantities, $\omega_a$ and $\omega_p$.  The term $\omega_a$ is the precession frequency measured with high energy decay positrons and $\omega_p$ is the magnetic field (B) normalized to the proton lamour frequency.  The spin and cyclotron frequencies can be combined to give the anomaly in terms of $\omega_a$, B, and the charge over mass ratio (e/m).  This can be rewritten in terms of $\omega_a$/$\omega_p$ and the muon, proton magnetic moment ratio from hyperfine splitting.   There is an additional electric field term which is removed by choosing the appropriate momenta of the muons, 3.09 GeV/c, known colloquially as the magic momentum.  Not all muons will have exactly this momentum, which introduces some uncertainties and corrections into the analysis.  Alignment efforts to ensure conformity of the muons are particularly important.

\subsection{Status of the Ring Installation}
\begin{figure}
\begin{center}
\includegraphics[width=100mm]{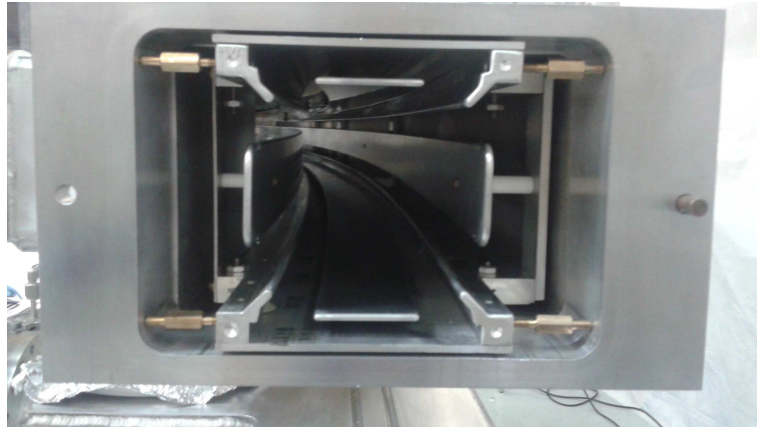}
\end{center}
\caption{\label{cage}View of the cage with trolley rails and (most interior) four electrostatic quadrapole plates, from a vacuum chamber end flange.}
\end{figure}

The vacuum chambers contain the electrostatic quadrapole plates, as mentioned earlier.  These plates are mounted to a cage structure which contains rails the B field measurement trolley will ride on, see Figure~\ref{cage}.  The quadrapole plates are aligned to the design value to within $\pm$0.5~mm (top/bottom) or $\pm$0.75~mm (sides) over the length of the chamber, with larger $\pm$2.0~mm deviations allowed on shorter length scales.  The cage rails are aligned to within $\pm$0.5~mm vertically to allow the trolley to ride smoothly, and the cage (and plates) is positioned radially to be centered on the magic radius, 7112~mm, the location where muons will have the magic momentum value.  The vertical alignment is done using a Hamar laser system along with a retroreflector, which allows a final measurement of the vertical alignment in vacuum, through a clear plexiglass flange.  Digital calipers and micrometers are used for quadrapole plate alignment, for absolute vertical alignment before the use of the relative laser system, and for the radial alignment.
\subsection{Status of the Magnetic Field Uniformity}
To produce a uniform magnetic field, iron shims are strategically added throughout the ring.  The field is measured and shims are added iteratively.  Measurements use a  shimming cart containing 25 nuclear magnetic resonance (NMR) probes and 4 capacitative gap sensors as well as 4 position sensors to give the cart location in r, $\theta$, and z.

\begin{figure}
\begin{center}
\includegraphics[scale=0.3]{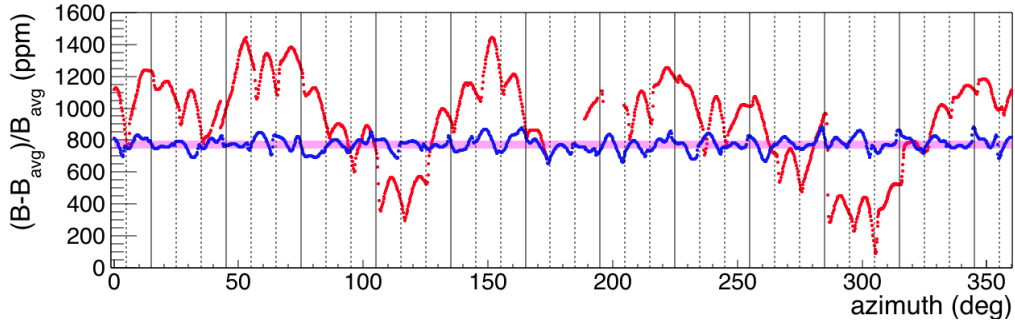}
\end{center}
\caption{\label{field1} Magnetic field variation versus azimuth for the first survey (red) and a more recent June 2016 survey (blue) with the variation goal shown in a magenta band.  Additional improvements to the field uniformity will continue through August 2016.}
\end{figure}

\begin{figure}
\begin{center}
\includegraphics[scale=0.45]{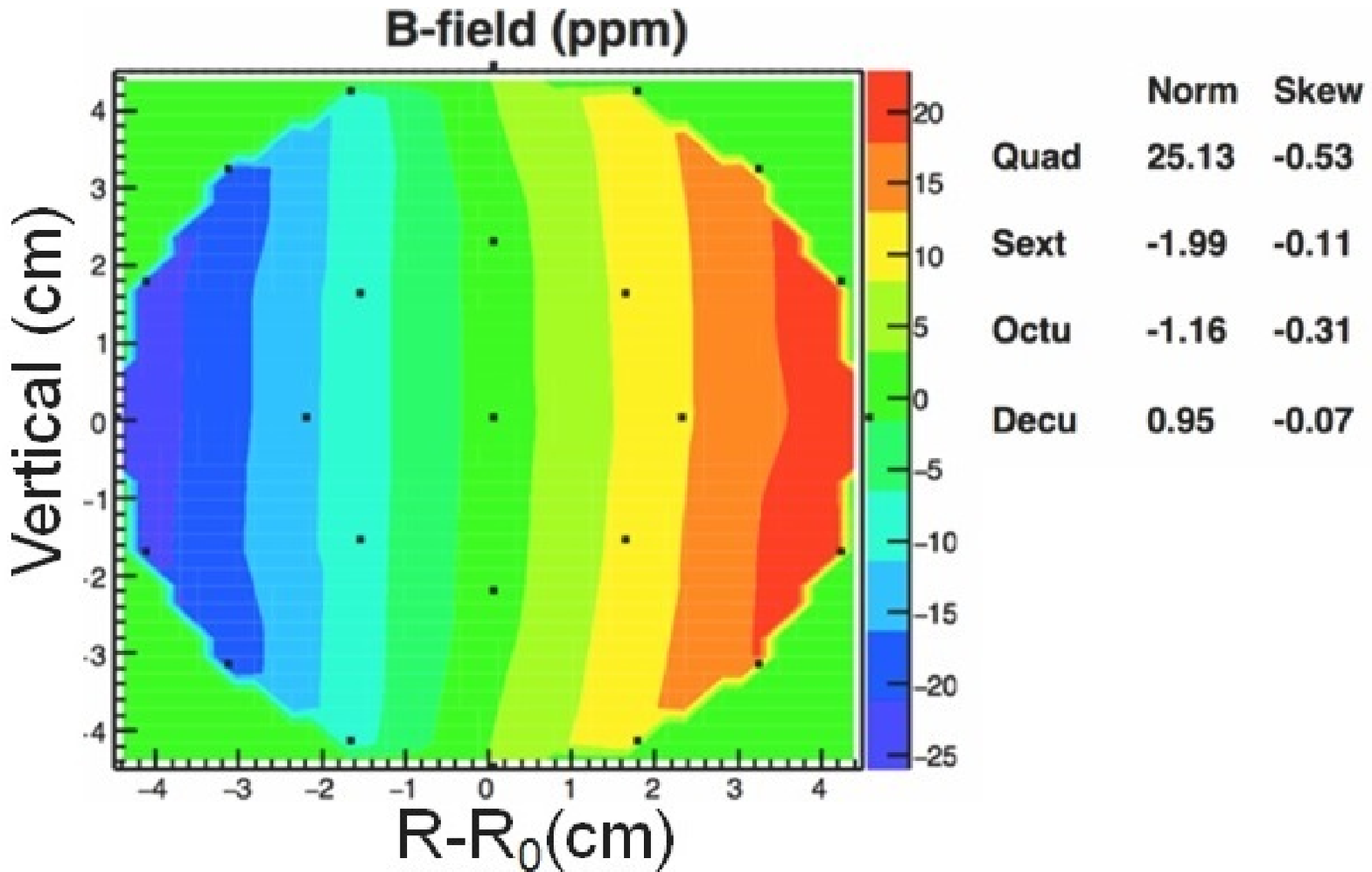}
\includegraphics[scale=0.45]{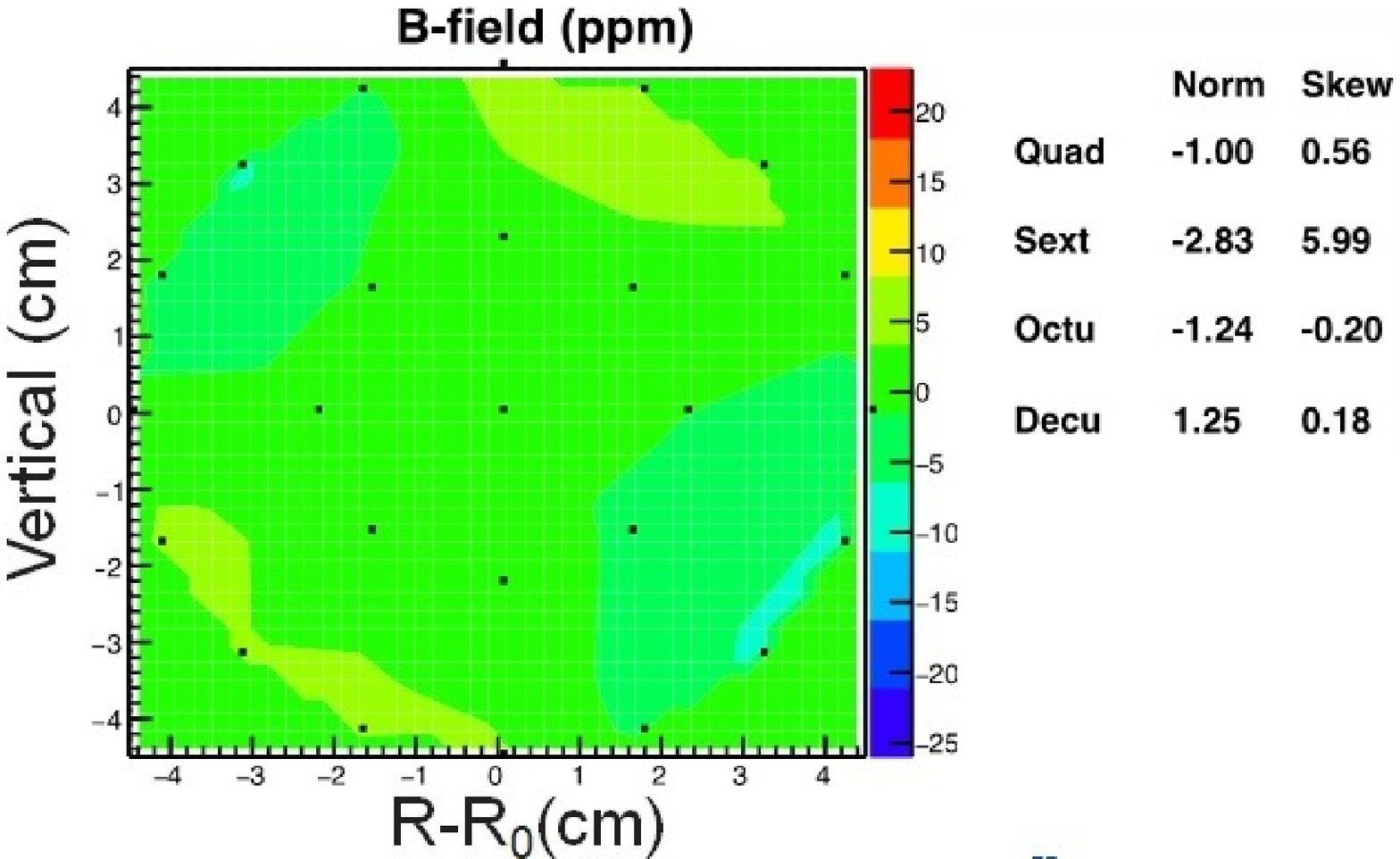}
\end{center}
\caption{\label{field2} Azmuthally averaged magnetic field variation for the first survey (left) and a more recent June 2016 survey (right).  Additional improvements to the field uniformity will continue through August 2016.}
\end{figure}

The initial field measurements when the magnet was first powered up were similar to those at BNL when it was first powered up.  A variation was found of $\pm$700~ppm in the field vs azimuth and $\pm$25~ppm in the azimuthally averaged field.  As of June 2016, the variation of the field has been reduced to $\pm$200~ppm (RMS of about 40~ppm), as shown in Figure~\ref{field1} and the azimuthally averaged field variation is reduced to  $\pm$6~ppm, as shown in Figure~\ref{field2}.  These are much closer to the target field variations of $\pm$25~ppm and $\pm <$~1~ppm, even before the planned laminated shims are applied.  This shim application is expected to finish in August of 2016.

\section{Summary}
The Muon g-2 Experiment at Fermilab is well under way in terms of its construction and assembly.  It is expected that the new experiment will improve the uncertainty on the measured value versus the previous measurement by a factor of four.  The collaboration is very active and working hard to install the various ring components to ensure a uniform B field and correct muon location.  Detector systems will be installed soon as well and data taking is expected in 2017, with the potential to resolve the tension of the previous result and the standard model calculation. 

\section*{References}

\bibliographystyle{iop-num}
\bibliography{holzbauer}

\end{document}